\documentclass[
 reprint,
 amsmath,amssymb,superscriptaddress,
 aps,
 prl,
]{revtex4-1}

\usepackage{graphicx}
\usepackage{color}
\usepackage{dcolumn}
\usepackage{bm}

\begin{document}

\title{Anisotropic Fermi Contour of (001) GaAs Electrons in Parallel Magnetic Fields}
\date{\today}

\author{D.\ Kamburov}
\author{M.~A.\ Mueed}
\author{M.\ Shayegan}
\author{L.~N.\ Pfeiffer}
\author{K.~W.\ West}
\author{K.~W.\ Baldwin}
\author{J.~J.~D.\ Lee}
\affiliation{ Department of Electrical Engineering, Princeton University, Princeton, New Jersey 08544, USA}
\author{R.\ Winkler}
\affiliation{Department of Physics, Northern Illinois University, DeKalb, Illinois 60115, USA}
\affiliation{Materials Science Division, Argonne National Laboratory, Argonne, Illinois 60439, USA}

\begin{abstract}
We demonstrate a severe Fermi contour anisotropy induced by the application of a parallel magnetic field to high-mobility electrons confined to a 30-nm-wide (001) GaAs quantum well. We study commensurability oscillations, namely geometrical resonances of the electron orbits with a unidirectional, surface-strain-induced, periodic potential modulation, to directly probe the size of the Fermi contours along and perpendicular to the parallel field. Their areas are obtained from the Shubnikov-de Haas oscillations. Our experimental data agree semi-quantitatively with the results of parameter-free calculations of the Fermi contours but there are significant discrepancies. 
\end{abstract}


\maketitle

An isotropic two-dimensional (2D) carrier system is characterized by a circular Fermi contour. In such a system, the application of a small perpendicular magnetic field leads to circular quasi-classical cyclotron orbits. If the layer of charged carriers is purely 2D, i.e., has zero thickness, the application of a parallel magnetic field ($B_\|$) would not affect the shape of its Fermi contour and the cyclotron trajectories would remain circular. However, if the layer has a finite (non-zero) thickness, $B_\|$ couples to the carriers' out-of-plane motion and distorts the Fermi contours and the cyclotron orbits \cite{Smrcka.JP.1993, Ohtsuka.PB.1998, Oto.PE.2001}. Understanding this $B_\|$-induced Fermi contour anisotropy is important for devices whose operation relies on ballistic transport \cite{Potok.PRL.2002}. The anisotropy also emerges in the context of magnetic breakdown and Fermi contour disintegration in bilayer systems \cite{Harff.PRB.1997, Jungwirth.PRB.1997}.

\begin{figure}[!b]
\includegraphics[trim=0.8cm 1cm 1cm 1.1cm, clip=true, width=0.49\textwidth]{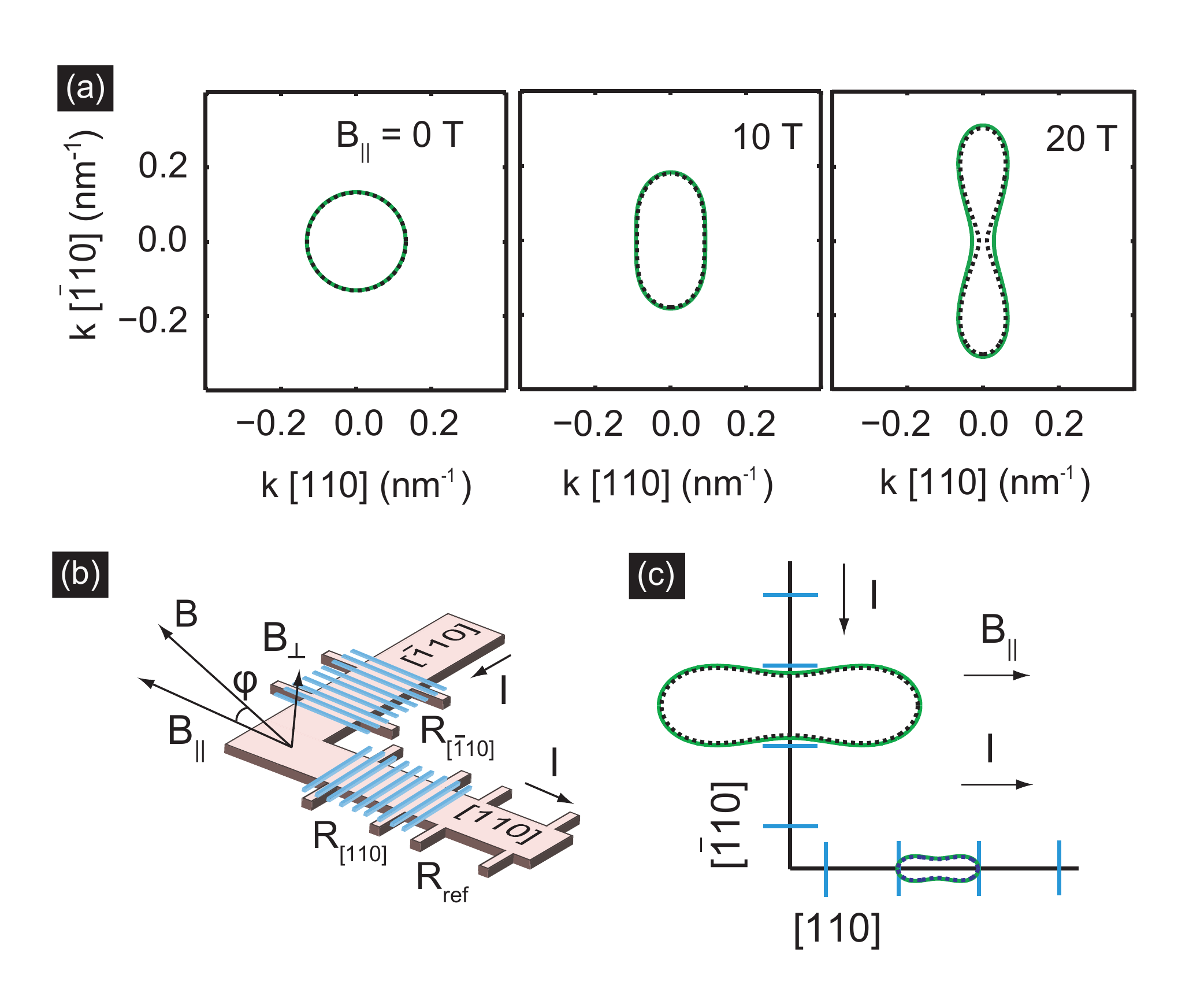}
\caption{\label{fig:Fig1} (color online) (a) Calculated 2D
electron Fermi contours for a 30-nm-wide GaAs quantum well when $B_\|$ is applied along the $[110]$ direction. The majority- (minority-) spin contour is given by solid (dotted) lines. (b) Schematic of the L-shaped Hall bar. The arms of the Hall bar, oriented along the $[110]$ and $[\overline{1}10]$ directions, are covered with stripes of negative electron-beam resist. Part of the Hall bar is intentionally left unpatterned and its magnetoresistance ($R_{\text{ref}}$) is used to measure Shubnikov-de Haas oscillations. (c) The geometry of the Hall bar is designed to use the commensurability of the ballistic cyclotron orbits with the period of the potential modulation induced by the stripes to probe the size of the Fermi wave vector along the $[110]$ and $[\overline{1}10]$ directions directly. Note that the real-space orbits are rotated by $90^{\circ}$ with respect to the Fermi contours \cite{Ashcroft}. }
\end{figure}

Here we demonstrate the ability to tune and measure the $B_\|$-induced Fermi contour anisotropy of electrons confined to a 30-nm-wide GaAs quantum well. Using geometrical resonances of cyclotron orbits with a periodic superlattice, the so-called commensurability oscillations (COs) \cite{Weiss.EP.1989, Winkler.PRL.1989, Gerhardts.PRL.1989, Beenakker.PRL.1989, Beton.PRB.1990, Peeters.PRB.1992, Mirlin.PRB.1998}, we directly probe the resulting distortions of the Fermi contour and the ballistic electron trajectories. Measuring Shubnikov-de Haas (SdH) oscillations allows us to determine the evolution of the Fermi contour areas with $B_\|$. Our results show that the Fermi contour distortion is significant and leads to a contour anisotropy of $\simeq 3.5 : 1$ for $B_\| \simeq 20$~T in our sample. This is much higher than the previously reported anisotropy in GaAs/AlGaAs heterojunctions \cite{Smrcka.JP.1993, Ohtsuka.PB.1998, Oto.PE.2001} and stems from the larger thickness of the electron layer in our sample. In contrast with the $B_{||}$-induced anisotropy in hole samples \cite{Kamburov.PRB.2012}, the electron anisotropy appears to be spin-independent. Comparison of our data with the results of numerical calculations reveals generally good agreement, although there are also significant disagreements.

Figure~\ref{fig:Fig1} captures the key points of our study. In Fig.~\ref{fig:Fig1}(a) we show the results of parameter-free calculations of the Fermi contours, combining the $8 \times 8$ Kane Hamiltonian \cite{RWinkler.book.2003} with spin-density functional theory \cite{Attaccalite.PRL.2002} to take into account the exchange-correlation of the quasi-2D electrons in our sample. At $B_\|=0$~T, the Fermi contours of the two spin-subbands are circular and essentially identical. With the application of $B_\|$ along the $[110]$ direction, both contours become elongated in the $[\overline{1}10]$ direction while shrinking along $[110]$. The areas enclosed by the two contours also differ from each other as electrons are transfered from the minority- to the majority-spin subbands. In our study we measure surface-strain-induced COs \cite{Skuras.APL.1997, Endo.PRB.2000, Endo.PRB.2001, Endo.PRB.2005, Kamburov.PRB._2012, Kamburov.PRB.2012}, triggered by a periodic density modulation [Figs.~\ref{fig:Fig1}(b) and (c)] to directly map the Fermi wave vectors in two perpendicular directions, $[\overline{1}10]$ and $[110]$.

\begin{figure}[t]
\includegraphics[trim=3.6cm 3cm 1cm 0.8cm, clip=true, width=0.48\textwidth]{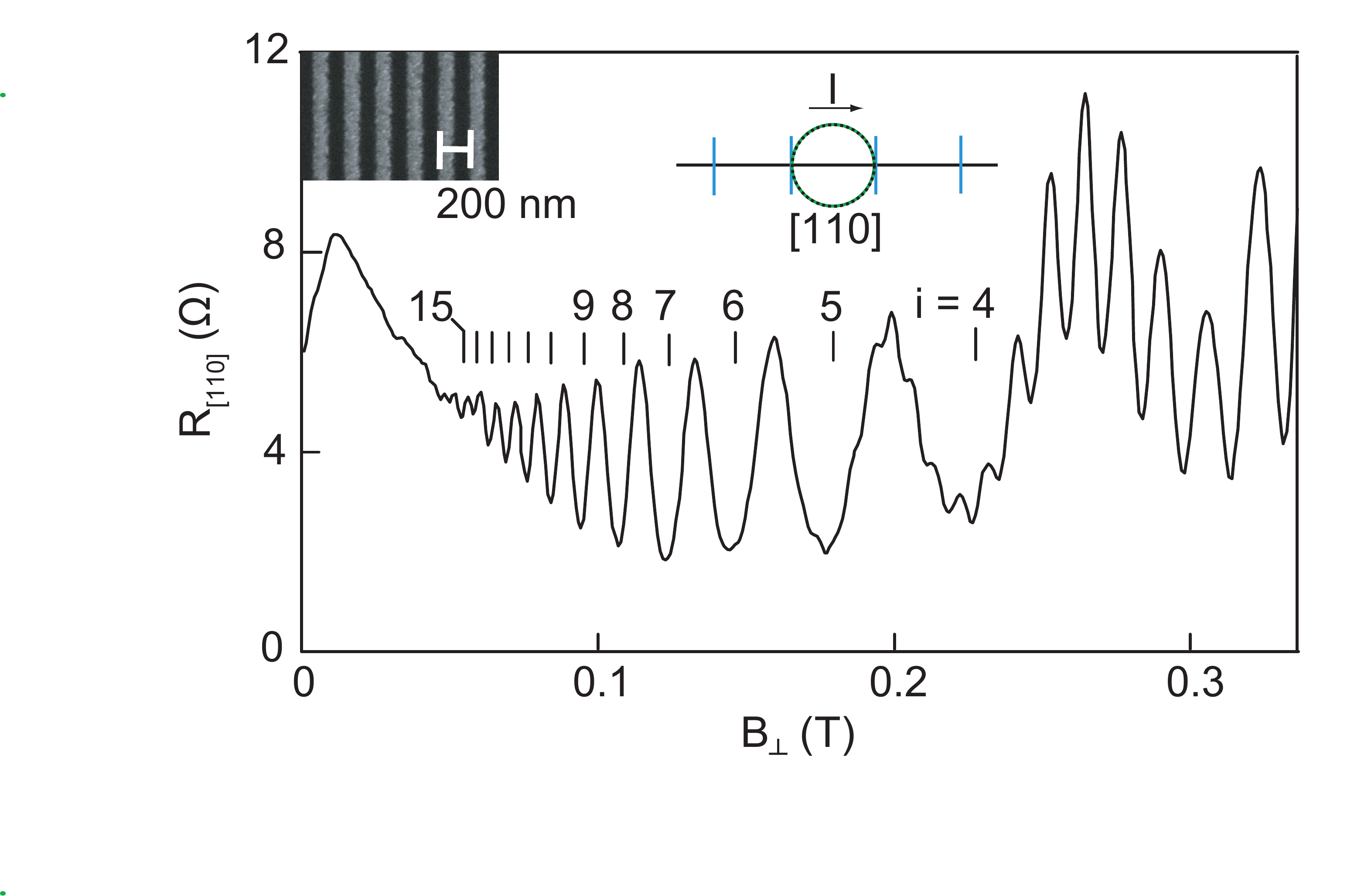}
\caption{\label{fig:Fig2} Low-field magnetoresistance measured in
the [110] direction at $B_\| = 0$ showing pronounced COs. Vertical lines mark the positions of the expected COs resistance minima according to $2R_{C}/a=i - 1/4$. Shubnikov-de Haas oscillations are visible on top of the COs above 0.2~T. The large number of COs minima (up to $\simeq 17$) attests to the very high quality of the sample and the periodic modulation. Inset: an example SEM image of the 200-nm-period grating of negative electron-beam resist.}
\end{figure}

The magnetoresistance of the modulated sections of our Hall bars exhibits minima at the electrostatic commensurability condition $2R_C/a=i-1/4,$ where $i=1,2,3, \ldots$ \cite{Weiss.EP.1989, Winkler.PRL.1989, Gerhardts.PRL.1989, Beenakker.PRL.1989, Beton.PRB.1990, Peeters.PRB.1992, Mirlin.PRB.1998}; an example is shown in Fig.~\ref{fig:Fig2}. Here $2R_C = 2k_F/eB$ is the real-space cyclotron diameter along the modulation direction and $a$ is the period of the potential modulation ($k_F$ is the Fermi wave-vector \textit{perpendicular} to the modulation direction) \cite{Gunawan.PRL.2004}. The anisotropy of the cyclotron diameter and the Fermi contour can therefore be quantified directly from COs measured along the two perpendicular arms of the L-shaped Hall bar in Fig.~\ref{fig:Fig1}(c). The COs for the arms along $[110]$ and $[\overline{1}10]$ yield $k_F$ along $[\overline{1}10]$ and $[110]$, respectively. In our measurements, we also recorded SdH oscillations in the unpatterned (reference) part of the Hall bar to probe the area enclosed by each of the Fermi contours.

\begin{figure}[t]
\includegraphics[trim=1cm 0.2cm 0.4cm 0.2cm, clip=true, width=0.48\textwidth]{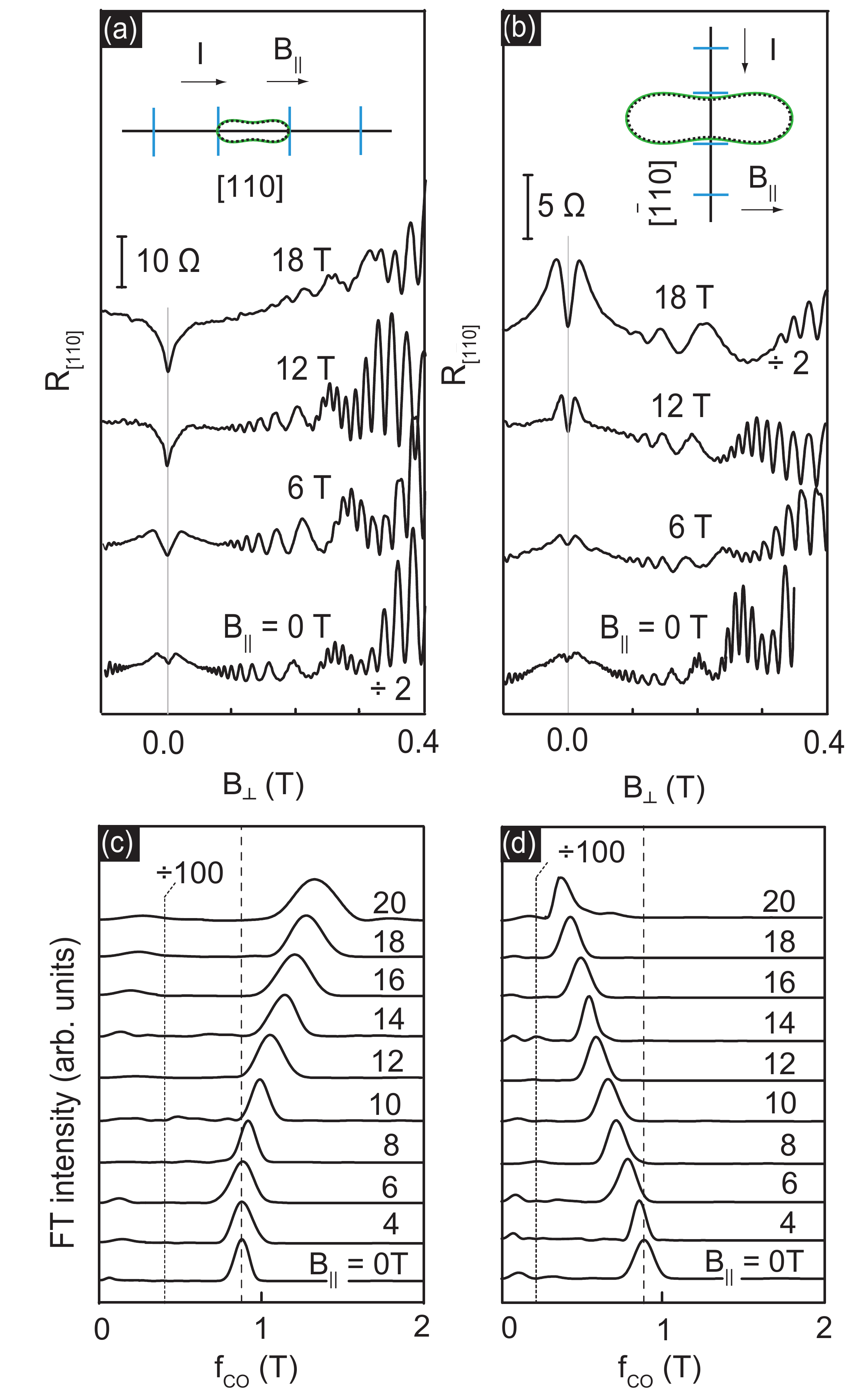}
\caption{\label{fig:Fig3} (color online) (a), (b)
Magnetoresistance data for the patterned sections of the L-shaped Hall bar in the $[110]$ and $[\overline{1}10]$ directions at different values of $B_\|$. (c), (d) Normalized Fourier transform spectra of the COs data shown in (a) and (b), respectively. The anticipated $B_\|{=}0$ COs frequency, based on a spin-degenerate, circular Fermi contour, is marked with dashed lines. The low-frequency parts of the spectra (below the vertical dotted lines) are severely affected by the Hamming window used in the Fourier analysis and are shown here suppressed by a factor of 100.}
\end{figure}

We prepared strain-induced superlattice samples with a lattice period of $a=200$ nm and 2D electrons confined to a 30-nm-wide GaAs quantum well grown via molecular beam epitaxy on a (001) GaAs substrate. The superlattice is made of negative electron-beam resist and modulates the 2D potential through the piezoelectric effect in GaAs \cite{Skuras.APL.1997, Endo.PRB.2000, Endo.PRB.2001, Endo.PRB.2005, Kamburov.PRB._2012, Kamburov.PRB.2012}. The quantum well, located 135 nm under the surface, is flanked on each side by 95-nm-thick Al$_{0.24}$Ga$_{0.76}$As spacer layers and Si $\delta$-doped layers. The 2D electron density at $T\simeq$ 0.3 K is $n\simeq 2.84 \times 10^{11}$ cm$^{-2}$, and the mobility is $\mu=18.4\times10^{6}$ cm$^2$/Vs. We passed current along the two Hall bar arms of the sample [Fig.~\ref{fig:Fig1}(b)] and measured the longitudinal resistances simultaneously along both arms. The measurements were carried out by first applying a fixed, large magnetic field in the plane of the sample along $[110]$. We then slowly rotated the sample around the [$\overline{1}10$] axis to introduce a small magnetic field ($B_{\perp}$) perpendicular to the 2D plane \cite{Tutuc.PRL.2001, footnote1}. This $B_{\perp}$ induced COs and SdH oscillations in our sample. The magnitude of $B_{\perp}$ was extracted from the Hall resistance we measured in the reference region of the sample simultaneously with the resistances of the two patterned regions. We performed all experiments using low-frequency ($\sim 13-18$ Hz) lock-in techniques in a $^3$He cryostat with a base temperature of $T\simeq 0.3$~K.

The magnetoresistance data from the two perpendicular Hall bar arms are shown in Figs.~\ref{fig:Fig3}(a) and (b). In each pannel the bottom traces, taken in the absence of $B_\|$, exhibit clear COs. The Fourier transform (FT) spectra of these two traces are shown as the bottom curves in Figs.~\ref{fig:Fig3}(c) and (d). Each of the FT spectra exhibits one peak whose position ($\simeq 0.88$~T) agrees with the commensurability frequency $f_{CO}=2 \hbar k_F/ea=0.88$~T expected for a circular, spin-degenerate Fermi contour with $k_F= \sqrt{2 \pi n}$ \cite{Weiss.EP.1989, Winkler.PRL.1989, Gerhardts.PRL.1989, Beenakker.PRL.1989, Beton.PRB.1990, Peeters.PRB.1992, Mirlin.PRB.1998}. With increasing $B_\|$, the peak in the FTs for the [110] Hall bar data [Fig.~\ref{fig:Fig3}(c)] moves to higher frequencies. In sharp contrast, the peak in the $[\overline{1}10]$ direction [Fig.~\ref{fig:Fig3}(d)] moves to smaller frequencies as $B_\|$ increases.

\begin{figure}[t]
\includegraphics[trim=2.2cm 0.6cm 0.6cm 2.6cm, clip=true, width=0.48\textwidth]{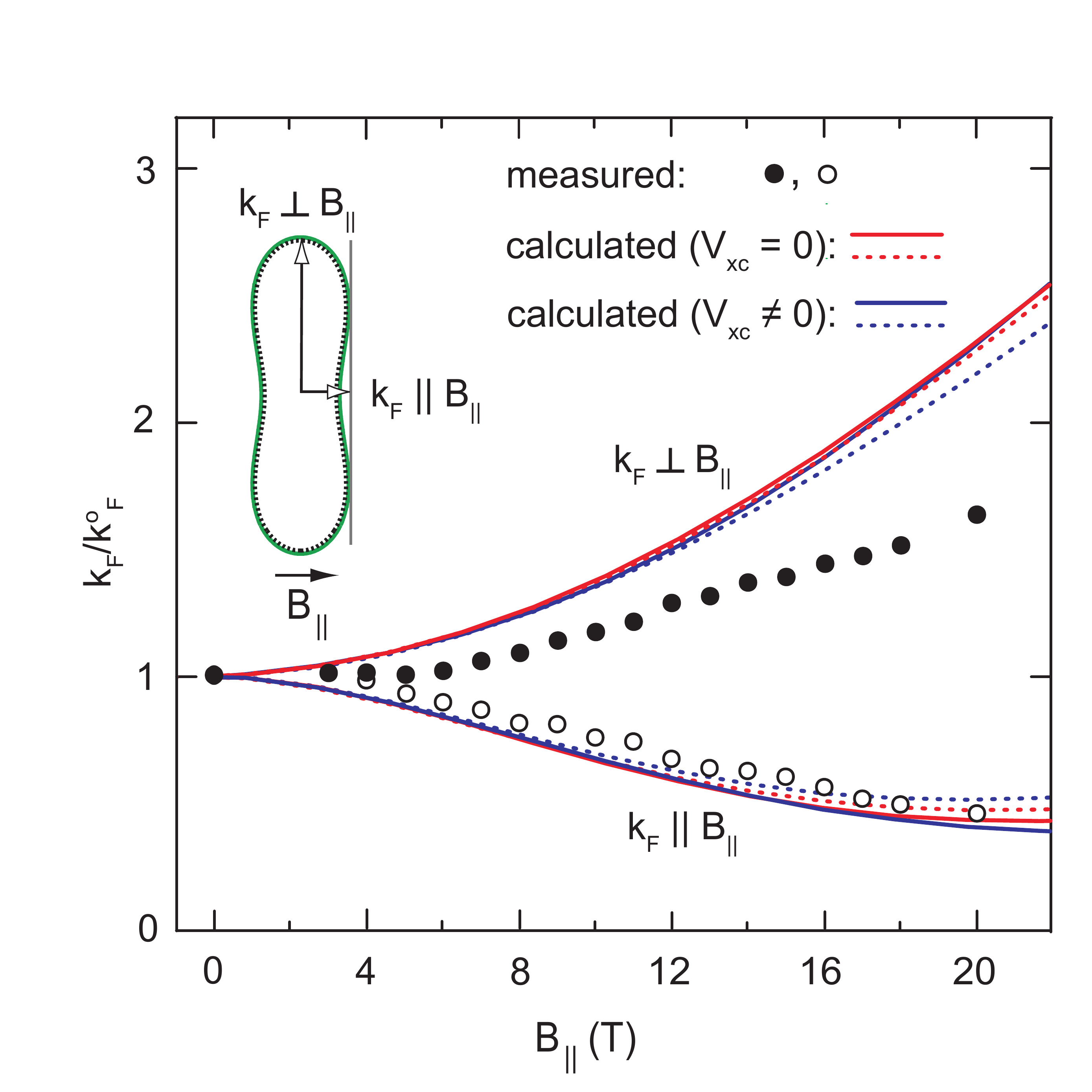}
\caption{\label{fig:Fig4} (color online) Summary of the Fermi wave
vector ($k_F$) peak deduced from the positions of the COs' FT spectra for the two Hall bar arms. Filled circles represent $k_F \perp B_\|$ and open circles the $k_F \parallel B_\|$ experimental data. Values of $k_F$ in the two directions calculated using $V_{xc} = 0$ and $V_{xc} \neq 0$ are given by red and blue lines, respectively. The calculated $k_F$ corresponding to the majority-spin species are plotted using solid lines, and the minority-spin species using dotted lines. }
\end{figure}

Figure~\ref{fig:Fig4} summarizes the measured $k_{F}$ as a function of $B_\|$, normalized to its value $k_F^{\circ}$ at $B_\|=0$. Similarly, we plot the extreme values of the Fermi wave vectors predicted by our parameter-free calculations using the $8 \times 8$ Kane Hamiltonian \cite{RWinkler.book.2003}. We include results from calculations that treat the exchange-correlation energy $(V_{xc})$ differently. The $V_{xc}=0$ calculation (red curves) ignores exchange-correlation completely while the $V_{xc} \neq 0$ calculation (blue curves) uses spin-density functional theory \cite{Attaccalite.PRL.2002} to take into account exchange-correlation in the 2D electron system that is partially spin-polarized because of $B_\|$.

The evolution of the COs' FT peaks with increasing $B_\|$ is qualitatively consistent with the calculated Fermi contours. The agreement is quantitatively good but for the $k_F \perp B$ case the elongation deduced from the experimental data is smaller than the calculations predict. This discrepancy implies that the shape of the Fermi contour is less elongated. We do not know the source of this disagreement at the moment. We note that we have encountered a similar disagreement in our study of \textit{hole} Fermi contours \cite{Kamburov.PRB.2012}. Despite this discrepancy, however, the overall agreement between the measured and calculated values of $k_{F}$ is remarkable, considering that there are no adjustable parameters in the calculations. The results of Fig.~\ref{fig:Fig4} clearly point to a severe distortion of the Fermi contours and the associated real-space ballistic electron trajectories in the presence of a moderately strong $B_\|$. Both calculations show that the extreme sizes of the contours for the two spin species remain very similar, explaining why the COs' FT peaks show no splittling \cite{footnote2}.

\begin{figure*}[t]
\includegraphics[trim=1.0cm 1.4cm 0.6cm 1.0cm, clip=true, width=0.90\textwidth]{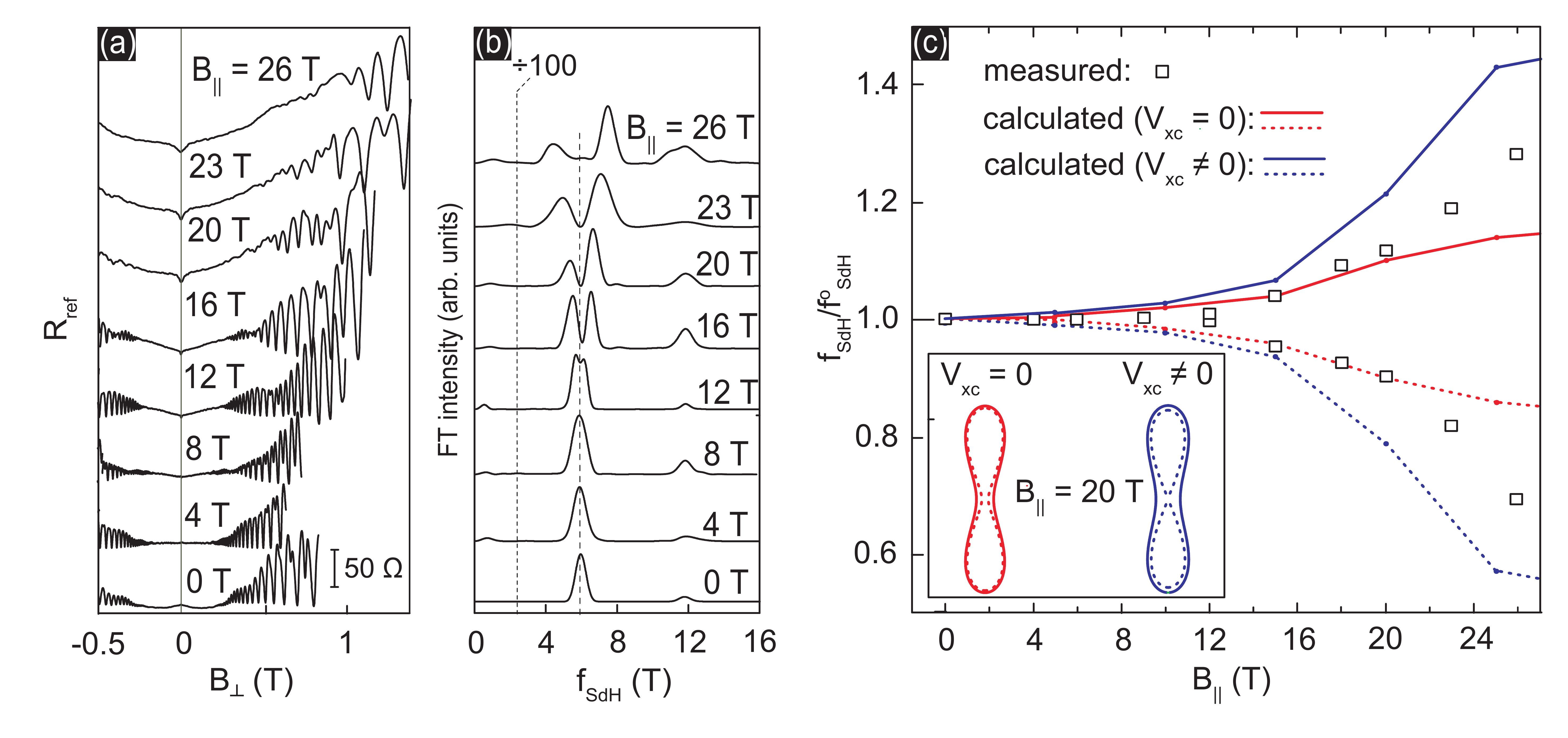}
\caption{\label{fig:Fig5} (color online) (a) Shubnikov-de Haas
oscillations measured in the unpatterned (reference) region of the Hall bar as $B_\|$ increases. (b) Fourier transform spectra of the SdH oscillations as a function of $B_\|$. The dashed line shows the expected position of the spin-unresolved FT peak. The signal in the region to the left of the vertical dotted line is shown suppressed. (c) Summary of the SdH FT peak positions normaized to $f^{\circ}_{SdH}$, the frequency at $B_\|=0$. Open squares represent the measured frequencies. The frequencies predicted by the calculations with $V_{xc} = 0$ and $V_{xc} \neq 0$ are shown using red and blue lines, respectively. For each calculation, the solid lines represent the majority-spin density, and the dotted lines the minority-spin density. Inset: Fermi contours corresponding to the $V_{xc} = 0$ (red) and $V_{xc} \neq 0$ (blue) calculations at $B_\|=20$~T.}
\end{figure*}

The COs data in Figs.~\ref{fig:Fig3} and \ref{fig:Fig4} probe the electron Fermi contours in two specific directions in $k$-space but give no information about their areas. To probe the areas enclosed by the Fermi contours, we measured the SdH oscillations in the unpatterned region of the sample [$R_{\text{ref}}$ in Fig.~\ref{fig:Fig1}(b)]. Figure~\ref{fig:Fig5}(a) shows the magnetoresistance traces at different $B_\|$. Their corresponding FTs are shown in Fig.~\ref{fig:Fig5}(b). Up to $B_\|=10$~T, the FT of each trace has two peaks. The position of the stronger peak is very close to the value of $(h/2e)n \simeq 5.8$~T expected for spin-unresolved SdH oscillations of electrons of density $n \simeq 2.8 \times 10^{11}$ cm$^{-2}$. The weaker peak at 11.6~T corresponds to spin-resolved oscillations [$(h/e)n=11.6$~T]. Starting at $B_\| \simeq 12$~T, the spin-unresolved peak at 5.8~T splits, with the upper peak corresponding to the area (electron density) of the majority-spin-subband and the lower peak to the minority-spin-subband.

Figure~\ref{fig:Fig5}(c) summarizes, as a function of $B_\|$, the measured SdH frequencies ($f_{SdH}$) normalized to the frequency $f^{\circ}_{SdH} \equiv 5.8$~T at $B_\|=0$, and the results of our energy band calculations. Overall, there is good qualitative agreement between the measured and calculated Fermi contour areas. Quantitatively, however, the experimental results fall between the calculated values with $V_{xc}=0$ and $V_{xc} \neq 0$. The differences between the two calculations are vizualized in the inset of Fig.~\ref{fig:Fig5}(c). When $V_{xc}=0$, the system is less spin-polarized and the areas enclosed by the Fermi contours of the two spin species are similar. When $V_{xc} \neq 0$, more charge is transferred from the minority- to the majority-spin species.

The experimental data and the numerical calculations presented here shed light on the shape of the electron Fermi contours in the presence of $B_\|$. The Fermi contour distortions implied by our data are by far larger than the distortions ($\simeq 10 \%$ at $B_\|=10$~T) expected or seen for 2D electrons confined to GaAs/AlGaAs heterojunctions \cite{Smrcka.JP.1993, Ohtsuka.PB.1998, Oto.PE.2001, Potok.PRL.2002}. This is mainly because of the larger thickness of the electron wave function in our 30-nm-wide quantum well sample. However, we emphasize that, besides the finite thickness of the carrier layer, other factors, such as the non-parabolicity of the energy bands and the spin-orbit interaction, also affect the distortion. For example, in 2D \textit{holes} confined to a much narrower 17.5-nm-wide quantum well, the distortions are yet larger than the ones reported here. At $B_\|=15$~T, the Fermi contour anisotropy there is $\simeq 3:1$ \cite{Kamburov.PRB.2012}, while the distortion we see here (Fig.~\ref{fig:Fig4}) is only $\simeq 1.6 : 1$. Furthermore, in contrast to the data presented here, the Fermi contour anisotropy exhibited by holes is very much spin-dependent: the majority-spin contour is much more elongated than the minority-spin contour. This strong spin-dependence stems from the much stronger spin-orbit interaction in 2D hole systems \cite{RWinkler.book.2003}. Finally, from the measured extremal $k_F$ (Fig.~\ref{fig:Fig4}), it appears that the Fermi contours are less elongated than the calculations predict. Remarkably, there is a similar discrepancy between the calculated and measured $k_F \perp B_\|$ for 2D \textit{hole} samples \cite{Kamburov.PRB.2012}.

\begin{acknowledgments}
We acknowledge support through the DOE BES (DE-FG02-00-ER45841) for measurements, and the Moore and Keck Foundations and the NSF (ECCS-1001719, DMR-1305691, and MRSEC DMR-0819860) for sample fabrication and characterization. A portion of this work was performed at the National High Magnetic Field Laboratory which is supported by National Science Foundation Cooperative Agreement No. DMR-1157490, the State of Florida and the US Department of Energy. Work at Argonne was supported by DOE BES under Contract No. DE-AC02-06CH11357. We thank S. Hannahs, T. Murphy, and A. Suslov at NHMFL for valuable technical support during the measurements. We also express gratitude to Tokoyama Corporation for supplying the negative e-beam resist \mbox{TEBN-1} used to make the samples.
\end{acknowledgments}


\begin{thebibliography}{99}

 

\bibitem{Smrcka.JP.1993} L. Smrcka and T. Jungwirth, J. Phys.: Condens. Matter \textbf{6}, 55 (1993).
\bibitem{Ohtsuka.PB.1998} K. Ohtsuka, S. Takaoka, K. Oto, K. Murase, and K. Gamo, Physica B \textbf{249}, 780 (1998).
\bibitem{Oto.PE.2001} K. Oto, S. Takaoka, K. Murase, and K. Gamo, Physica E \textbf{11}, 177 (2001).
\bibitem{Potok.PRL.2002} See, e.g., R. M. Potok, J. A. Folk, C. M. Marcus, and V. Umansky, Phys. Rev. Lett. \textbf{89}, 266602 (2002). In transverse magnetic focusing experiments reported in this work, an anomalous shift of the focusing peaks' positions is seen in Fig. 1 when $B_\|=7$ T. This shift, which is not discussed by Potok \textit{et al.}, stems from the distortion of the Fermi contour with $B_\|$.
\bibitem{Harff.PRB.1997} N. E. Harff, J. A. Simmons, J. F. Klem, G. S. Boebinger, L. N. Pfeiffer, and K. W. West, Superlattices and Microstructures \textbf{20}, 595 (1996).
\bibitem{Jungwirth.PRB.1997} T. Jungwirth, T. S. Lay, L. Smrcka, and M. Shayegan, Phys. Rev. B \textbf{56}, 1029 (1997).
\bibitem{Weiss.EP.1989} D. Weiss, K. von Klitzing, K. Ploog, and G. Weimann, Europhys. Lett. \textbf{8}, 179 (1989).
\bibitem{Winkler.PRL.1989} R. W. Winkler, J.P. Kotthaus, and K. Ploog, Phys. Rev. Lett. \textbf{62}, 1177 (1989).
\bibitem{Gerhardts.PRL.1989} R. R. Gerhardts, D. Weiss, and K. von Klitzing, Phys. Rev. Lett. \textbf{62}, 1173 (1989).
\bibitem{Beenakker.PRL.1989} C. W. J. Beenakker, Phys. Rev. Lett. \textbf{62}, 2020 (1989).
\bibitem{Beton.PRB.1990} P. H. Beton, E. S. Alves, P. C. Main, L. Eaves, M. W. Dellow, M. Henini, O. H. Hughes, S. P. Beaumont, and C. D. W. Wilkinson, Phys. Rev. B \textbf{42}, 9229 (1990).
\bibitem{Peeters.PRB.1992} F. M. Peeters and P. Vasilopoulos, Phys. Rev. B \textbf{46}, 4667 (1992).
\bibitem{Mirlin.PRB.1998} A. D. Mirlin and P. Wolfle, Phys. Rev. B \textbf{58}, 12986 (1998).
\bibitem{Kamburov.PRB.2012} D. Kamburov, M. Shayegan, R. Winkler, L. N. Pfeiffer, K. W. West, and K. W. Baldwin, Phys. Rev. B \textbf{86}, 241302 (2012).
\bibitem{Ashcroft} N.W. Ashcroft and N.D. Mermin,\textit{ Solid State Physics} (Holt, Rinehart and Winston, Philadelphia, 1976), Chapter 12.
\bibitem{RWinkler.book.2003} R. Winkler, \textit{Spin-Orbit Coupling Effects in Two-Dimensional Electron and Hole Systems} (Springer, Berlin, 2003).
\bibitem{Attaccalite.PRL.2002} C. Attaccalite, S. Moroni, P. Gori-Giori, and G. B. Bachelet, Phys. Rev. Lett. \textbf{88}, 256601 (2002).
\bibitem{Skuras.APL.1997} E. Skuras, A. R. Long, I. A. Larkin, J. H. Davies, and M. C. Holland, Appl. Phys. Lett. \textbf{70}, 871 (1997).
\bibitem{Endo.PRB.2000} A. Endo, S. Katsumoto, and Y. Iye, Phys. Rev. B \textbf{62}, 16761 (2000).
\bibitem{Endo.PRB.2001} A. Endo, M. Kawamura, S. Katsumoto, and Y. Iye, Phys. Rev. B \textbf{63}, 113310 (2001).
\bibitem{Endo.PRB.2005} A. Endo and Y. Iye, Phys. Rev. B \textbf{72}, 235303 (2005).
\bibitem{Kamburov.PRB._2012} D. Kamburov, H. Shapourian, M. Shayegan, L. N. Pfeiffer, K. W. West, K. W. Baldwin, and R. Winkler, Phys. Rev. B \textbf{85}, 121305(R) (2012).
\bibitem{Gunawan.PRL.2004} O. Gunawan, Y. P. Shkolnikov, E. P. De Poortere, E. Tutuc, and M. Shayegan, Phys. Rev. Lett. \textbf{93}, 246603 (2004).
\bibitem{Tutuc.PRL.2001} E. Tutuc, E. P. De Poortere, S. J. Papadakis, and M. Shayegan, Phys. Rev. Lett. \textbf{86}, 2858 (2001).
\bibitem{footnote1}  Note that, when the total applied field $B$ is large compared to $B_{\perp}$, the parallel component of the field, $B_\|=\sqrt{B^2-B_{\perp}^2}$, remains essentially fixed and equal to $B$ as we rotate the sample and take data.

\bibitem{footnote2} In Fig.~\ref{fig:Fig4}, for the calculated $k_F \parallel B_\|$ values, we have plotted $k_F$ corresponding to the maximum diameter of the Fermi contour along $B_\|$ (see the small, open-headed, horizontal arrow in Fig.~\ref{fig:Fig4} inset). The peanut-shape Fermi contour shown in the inset also has a slightly smaller maximal diameter in the same direction, corresponding to the ``neck'' of the peanut. In principle, one might expect COs associated with this $k_F$ also, but in our data [Fig.~\ref{fig:Fig3}(d)] we observe only one frequency for the COs.



\end{thebibliography}
\end{document}